\documentclass[preprint,aps,pra,showpacs,floatfix]{revtex4}
\usepackage{graphicx}
\usepackage{times}
\usepackage{nicefrac}
\usepackage{amsmath}
\usepackage{amsfonts}
\usepackage{amssymb}
\usepackage{amsthm}
\usepackage{epsf}
\usepackage{bm}
\usepackage{bbm}
\usepackage{times}

\usepackage{dcolumn}
\newcolumntype{.}{D{x}{}{-1}}

\newcommand{\be}{\begin{eqnarray}}
\newcommand{\ee}{\end{eqnarray}}

\newcommand{\veps}{\varepsilon}

\newcommand{\balpha}{\bm{\alpha}}

\newcommand{\bfr}{{\bf r}}
\newcommand{\bfx}{{\bf x}}

\newcommand{\ket}[1]{|#1\rangle}
\newcommand{\matrixel}[3]{\langle#1|#2|#3\rangle}
\newcommand{\az}{\alpha Z}

\newcommand{\gdirac}{g_{\rm D}}
\newcommand{\dgint}{\Delta g_{\rm int}}
\newcommand{\dgqed}{\Delta g_{\rm QED}}
\newcommand{\dgsqed}{\Delta g_{\rm SQED}}
\newcommand{\dgnuc}{\Delta g_{\rm nuc}}

\newcommand{\dgse}{\Delta g_{\rm SE}}
\newcommand{\dgvp}{\Delta g_{\rm VP}}

\newcommand{\dgirr}{\Delta g_{\rm irr}}
\newcommand{\dgred}{\Delta g_{\rm red}}
\newcommand{\dgver}{\Delta g_{\rm ver}}
\newcommand{\dgvr}{\Delta g_{\rm vr}}

\newcommand{\Vnucl}{V_{\rm nuc}}

\newcommand{\Veff}{V_{\rm eff}}

\newcommand{\xalpha}{x_{\alpha}}
\begin{document}

\title{
Screened QED corrections to the $\bm{g}$ factor of Li-like ions}
\author{
D.~A.~Glazov,$^{1}$ A.~V.~Volotka,$^{1,2}$ V.~M.~Shabaev,$^{1}$
I.~I.~Tupitsyn,$^{1}$ and G.~Plunien$^{2}$ }
\affiliation{
$^1$
Department of Physics, St. Petersburg State University,
Oulianovskaya 1, Petrodvorets, St. Petersburg 198504, Russia \\
$^2$ Institut f\"ur Theoretische Physik, TU Dresden,
Mommsenstra{\ss}e 13, D-01062 Dresden, Germany \\
}
\begin{abstract}
The screened QED corrections to the $g$ factor of Li-like ions
are evaluated utilizing an effective potential approach.
The evaluation is performed within the range of nuclear charge
numbers $Z=32\,$--$\,92$.
The results obtained serve for improving the theoretical predictions
for the $g$ factor of heavy Li-like ions.
\end{abstract}


\maketitle

%

High-precision measurements of the $g$ factor of H-like carbon and oxygen
\cite{hermanspahn:00:prl,haeffner:00:prl,verdu:04:prl,stahl:05:jpb}
stimulated accurate theoretical calculations of this effect
\cite{beier:00:pra,martynenko:01,
karshenboim:02:plb,nefiodov:02:prl,shabaev:01:pra,
shabaev:02:prl,yerokhin:04:pra,pachucki:05:pra,lee:05:pra}.
In particular, these investigations provided a new determination of the electron mass
to an accuracy which is four times better than that of the previously accepted value
(see Ref. \cite{CODATA02} and references therein).
Moreover, an extension of these studies to higher $Z$ systems could lead to an independent
determination of the fine structure constant
\cite{werth:01:hydr,karshenboim:01:hydr}.
Investigation of ions with more than one electron is also anticipated
in the nearest future. In particular, measurements of the $g$ factor
of Li-like calcium are currently in progress by the Mainz-GSI collaboration.

The motivation for studying Li-like ions alongside with H-like ions
arises from the substantial elimination of the uncertainty
due to the nuclear size effects in a specific combination
of the corresponding $g$ factor values \cite{shabaev:02:pra}.
The most accurate results for different contributions and for total values
of the $g$ factor of Li-like ions within the range $Z=6\,$--$\,92$ were presented
in our recent work \cite{glazov:04:pra}. For all values of $Z$
the uncertainty of the total value was mainly defined by the contributions
of the interelectronic interaction and the QED screening effect.
The latter was evaluated to the leading order $(\alpha/\pi)(\az)^2/Z$
and calculations to all orders in $\az$ are needed for high values of $Z$.
In the present paper we perform this evaluation using an effective potential approach.
The obtained results are combined with other contributions to improve theoretical
predictions for the $g$ factor of Li-like ions within the range $Z=32\,$--$\,92$.
Compared to our previous work \cite{glazov:04:pra},
the new theoretical values include also recent results for the higher-order
two-loop QED corrections \cite{pachucki:05:pra} and the magnetic-loop part
of the vacuum polarization \cite{lee:05:pra}. In addition, the nuclear-size
correction is recalculated employing the most recent data for nuclear radii from Ref.
\cite{angeli:04:adndt}.

The relativistic units ($\hbar=c=m=1$) and the Heaviside charge unit
($\alpha=e^2/(4\pi), e<0$) are used throughout the paper.


The total value of the ground-state $g$ factor of a Li-like ion is conveniently
written as
\be
\label{eq:gtotal}
  g = \gdirac + \dgint + \dgqed + \dgsqed + \dgnuc\,,
\ee
where
\be
  \gdirac = \frac23\,\left(1+\sqrt{2+2\sqrt{1-(\az)^2}}\right)
  = 2 - \frac{(\az)^2}{6} + \dots
\ee
is the Dirac value for the point-charge nucleus,
$\dgint$ is the interelectronic-interaction correction,
$\dgqed$ is the one-electron QED contribution,
$\dgsqed$ is the screened QED correction
and $\dgnuc$ incorporates the nuclear-size, nuclear-recoil
and nuclear-polarization corrections.
For the evaluation of the interelectronic-interaction and nuclear corrections
we refer to our previous papers
\cite{shabaev:02:pra,glazov:04:pra}
and references therein.
Radiative corrections for electrons bound in a pure Coulomb field were addressed,
in particular, in Refs.
\cite{beier:00:pra,yerokhin:04:pra,
pachucki:05:pra,karshenboim:02:plb,lee:05:pra}.
The main goal of the present work is to evaluate the screened QED corrections
within an effective potential approach.

We consider an effective spherically symmetric potential $\Veff$
that partly takes into account the interelectronic interaction
between the valence $2s$ electron and the core electrons of the $(1s)^2$ shell.
The simplest choice of $\Veff$ is the core-Hartree (CH) potential
\be
\label{eq:Vscr-cH}
  \Veff(r) = \Vnucl(r) + {\alpha} \int_0^{\infty}dr' \frac{1}{r_>} \rho_c(r'),
\ee
where $\rho_c$ is the density of the core electrons
and $\Vnucl$ denotes the nuclear potential.
The screening potential derived from the density-functional theory reads
\cite{cowan,indelicato:90:pra,sapirstein:02:pra}
\be
\label{eq:Vscr-dft}
  \Veff(r) = \Vnucl(r) + {\alpha} \int_0^{\infty}dr' \frac{1}{r_>} \rho_t(r')
  - \xalpha\,\frac{\alpha}{r} \left( \frac{81}{32\pi^2} r \rho_t(r) \right)^{1/3}.
\ee
Here $\rho_t$ is the total electron density, including the $(1s)^2$ shell
and the $2s$ electron. The parameter $\xalpha$ can be varied from $0$ to $1$.
The cases of values $\xalpha=0$, $2/3$ and $1$ are referred to as the Dirac-Hartree (DH),
the Kohn-Sham (KS) and the Dirac-Slater (DS) potentials, respectively.
To provide a proper asymptotic behavior, equation (\ref{eq:Vscr-dft})
should be replaced by \cite{latter}
\be
  \Veff(r) = -\frac{\alpha(Z-N_c)}{r}
\ee
at large $r$, where $N_c$ is the number of core electrons.
The self-consistent potential (\ref{eq:Vscr-dft}) is generated by iterations,
which continue until the energies of the core and valence states become stable
on the level of $10^{-9}$. The CH potential (\ref{eq:Vscr-cH}) does not imply
self-consistency and, therefore, it is generated after one iteration.
When the effective potential $\Veff$ and the spectrum of the corresponding
Dirac equation are generated, the screening correction $\dgsqed$ is calculated
as the difference between the two values of the QED correction calculated
for the potential $\Veff$ and for the nuclear potential $\Vnucl$,
\be
\label{eq:dgsqed}
  \dgsqed = \dgqed\left[\Veff\right] - \dgqed\left[\Vnucl\right] \,.
\ee

Below, we describe the evaluation of the one-loop QED correction to the $g$ factor
in arbitrary binding potential. The QED correction of first order in $\alpha$
appears as the sum of self-energy and vacuum-polarization corrections,
$\dgqed^{(1)} = \dgse + \dgvp$.
The vacuum-polarization term is relatively small, and its contribution
to the QED screening can be neglected at the present level of accuracy.
The self-energy correction is given by the sum of irreducible,
reducible and vertex parts,
\be
  \dgse = \dgirr + \dgred + \dgver \,.
\ee
The expression for the irreducible part reads \cite{ttgf}
\be
\label{eq:irr0}
  \dgirr = \frac{1}{m_a} \sum_{n}^{\veps_{n}\neq\veps_a}
  \frac{ \matrixel{a}{\left(\Sigma(\veps_a)-\gamma^0\delta m\right)}{n}
  \matrixel{n}{[\bfr\times\balpha]_z}{a} }{\veps_a-\veps_{n}} \,.
\ee
Here $\ket{a}$ is the $2s$ state with the angular momentum projection $m_a$,
$\delta m$ is the mass counter-term and $\Sigma(\veps)$ denotes
the unrenormalized self-energy operator defined by
\be
  \matrixel{a}{\Sigma(\veps)}{b} =
  \frac{i}{2\pi}\int_{-\infty}^{\infty} d\omega \sum_{n}
  \frac{\matrixel{a n}{I(\omega)}{n b}}{\veps-\omega-\veps_{n}(1-i0)}
  \,,
\ee
where
$I(\omega,\bfx_1,\bfx_2)=e^2\alpha^{\mu}\alpha^{\nu}D_{\mu\nu}(\omega,\bfx_1-\bfx_2)$,
$\alpha^{\mu}=(1,\balpha)$ are the Dirac matrices
and $D_{\mu\nu}$ is the photon propagator.
To separate ultraviolet divergent zero-potential $\dgirr^{(0)}$
and one-potential $\dgirr^{(1)}$ terms we follow
Ref. \cite{snyderman:91:ann}
and calculate them in momentum space.
The residual part of $\dgirr$, so-called many-potential term $\dgirr^{(2+)}$,
is calculated in coordinate space employing the algorithm
proposed in Ref. \cite{blundell:91:pra}.
The expression for the reducible part reads \cite{ttgf}
\be
\label{eq:red}
  \dgred = \frac{1}{m_a} \matrixel{a}{\frac{\partial}{\partial\veps}
  {\Sigma(\veps)\vrule}_{\veps=\veps_a}}{a} \matrixel{a}{[\bfr\times\balpha]_z}{a}\,,
\ee
while the vertex part is given by \cite{ttgf}
\be
\label{eq:ver}
  \dgver = \frac{1}{m_a} \frac{i}{2\pi} \int_{-\infty}^{\infty} d\omega
  \sum_{n_1,n_2} \frac{ \matrixel{a n_2}{I(\omega)}{n_1 a}
  \matrixel{n_1}{[\bfr\times\balpha]_z}{n_2} }
  {(\veps_a-\omega-\veps_{n_1}(1-i0))(\veps_a-\omega-\veps_{n_2}(1-i0))}\,.
\ee
Both reducible and vertex parts are ultraviolet divergent, whereas the sum
$\dgvr = \dgver + \dgred$ is finite.
Following Refs. \cite{beier:00:pra,yerokhin:04:pra},
we separate out the zero-potential term $\dgvr^{(0)}$ and evaluate it in momentum space.
The remaining many-potential term $\dgvr^{(1+)}$ is calculated in
coordinate space as the point-by-point difference between the contributions (\ref{eq:ver})
with bound and free propagators in the self-energy loop.

Below we discuss some details, concerning coordinate-space calculations
of $\dgirr^{(2+)}$ and $\dgvr^{(1+)}$. Angular integration and summation
over intermediate angular momentum projections is carried out in the standard way.
The many-potential terms involve
infinite summation over the angular quantum number $\kappa$.
The summation over the complete spectrum of the Dirac equation at fixed $\kappa$
is performed using the dual-kinetic-balance approach \cite{shabaev:04:prl}.
The summation over $\kappa$ is terminated at $|\kappa|=10\,$--$\,20$
and the rest of the sum is estimated by the least-square inverse-polynomial fitting.
It was observed that more stable values are obtained, when subtraction (\ref{eq:dgsqed})
is performed prior to the fitting procedure.

%

The results obtained for the screened QED correction are presented in Table~\ref{tab:sqed}.
The uncertainty of the numerical evaluation is defined by the many-potential terms.
The uncertainty due to incompleteness of the effective potential approach
in the description of the interelectronic interaction is, however, much larger.
It was estimated as the mean deviation of the results obtained by means of the different
potentials: core-Hartree, Kohn-Sham, Dirac-Hartree and Dirac-Slater.
Since for middle-$Z$ ions the accuracy of our previous results \cite{glazov:04:pra}
turns out to be better than that of the present ones, in the final compilation (see below)
for $Z<30$ we present the values of $\dgsqed$ from \cite{glazov:04:pra}.
We mention that these values include terms of higher order in $1/Z$
calculated by Yan \cite{yan:01:prl}.
For $Z>30$ we take the present data, obtained with the Kohn-Sham potential.
While for middle $Z$ there is a good agreement between the present
and the previous results, a significant discrepancy is found for high values of $Z$.
This is presumably due to the fact that the method of Ref. \cite{glazov:04:pra},
based on the non-relativistic treatment of the anomalous magnetic moment,
is rather insensitive to the bound-state QED effects. The accuracy of the present
evaluation of the screened QED correction is limited by the possibility to account for
the interelectronic-interaction effects in terms of the local potential $\Veff$.
A rigorous evaluation of the $1/Z$ contribution to $\dgsqed$ should be the next step
for improvement of the accuracy of the many-electron QED correction.

In Table~\ref{tab:total}, the individual contributions to the $g$ factor
of the ground state of Li-like ions are presented. The changes made
compared with the previous compilation \cite{glazov:04:pra} concern several terms:
the one-electron QED correction, the screened QED correction
and the finite-nuclear-size correction.
The evaluation of $\dgsqed$ is already described above.
The finite-nuclear-size correction is recalculated with the most recent data
for the nuclear radii \cite{angeli:04:adndt}.
The one-electron QED correction of first order in $\alpha$ is updated
with the recent evaluation of the magnetic-loop vacuum-polarization term
\cite{lee:05:pra}.
This reduces the uncertainty of the $\dgqed^{(1)}$ term for $Z>30$.
The one-electron QED correction of second order in $\alpha$ is improved
for $Z<30$ employing the analytical formula for the $(\alpha/\pi)^2(\az)^4$ term
recently derived in Ref. \cite{pachucki:05:pra}.

In summary, we have presented the evaluation of the screened QED correction
to the $g$ factor of Li-like ions within the effective potential approach.
These results improve the accuracy of the theoretical predictions
for the $g$ factor within the range $Z=32\,$--$\,92$,
where stringent tests of the bound-state QED effects are expected.
More elaborated treatment of the interelectronic-interaction and screened QED
corrections will be the subject of our subsequent investigations.

\acknowledgments
We thank A.~N.~Artemyev, N.~S.~Oreshkina, D.~A.~Solovyev and V.~A.~Yerokhin
for valuable conversations.
This work was supported by RFBR (Grant No. 04-02-17574)
and by INTAS-GSI (Grant No. 03-54-3604).
D.A.G. acknowledges the support by the ``Dynasty'' foundation.
A.V.V. and G.P. acknowledge financial support from the GSI F+E program, DFG, and BMBF.
%
%
%

%
%
%
\begin{table}
\caption{The screened QED correction $\dgsqed$ obtained with
core-Hartree (Eq. (\ref{eq:Vscr-cH})),
Kohn-Sham (Eq. (\ref{eq:Vscr-dft}), $\xalpha=2/3$),
Dirac-Hartree (Eq. (\ref{eq:Vscr-dft}), $\xalpha=0$)
and Dirac-Slater (Eq. (\ref{eq:Vscr-dft}), $\xalpha=1$) potentials.
The results of our previous evaluation \cite{glazov:04:pra}
are shown for comparison. All numbers are in units of $10^{-6}$.
\label{tab:sqed}}
\vspace{0.5cm}
\begin{tabular}{rr@{}lr@{}lr@{}lr@{}lr@{}l}
%
\hline
\hline
$Z$ &
\multicolumn{2}{c}{CH} &
\multicolumn{2}{c}{KS} &
\multicolumn{2}{c}{DH} &
\multicolumn{2}{c}{DS} &
\multicolumn{2}{c}{\cite{glazov:04:pra}} \\
%
\hline

18 &
$-$0.&322  &
$-$0.&217  &
$-$0.&235  &
$-$0.&180  &
$-$0.&292 (81) \\

20 &
$-$0.&371  &
$-$0.&244  &
$-$0.&260  &
$-$0.&205  &
$-$0.&33 (10)  \\

24 &
$-$0.&464  &
$-$0.&318  &
$-$0.&333  &
$-$0.&278  &
$-$0.&42 (15) \\

32 &
$-$0.&629  &
$-$0.&452  &
$-$0.&462  &
$-$0.&411  &
$-$0.&62 (27) \\

54 &
$-$1.&354  &
$-$1.&180  &
$-$1.&150  &
$-$1.&029  &
$-$1.&6 (8) \\

82 &
$-$5.&028  &
$-$3.&524  &
$-$3.&288  &
$-$3.&001  &
$-$5.&6 (2.0) \\

92 &
$-$4.&531  &
$-$3.&127  &
$-$2.&961  &
$-$2.&859  &
$-$9.&2 (2.6) \\

\hline
\hline
\end{tabular}
\end{table}
%
%
%
\squeezetable
\begin{table}
\caption{Individual contributions to the ground-state $g$ factor of Li-like ions.
\label{tab:total}}
\vspace{0.5cm}
\begin{tabular}{lr@{}lr@{}lr@{}lr@{}l}
\hline
\hline
& \multicolumn{2}{c}{$^{12}{\rm C}^{3+}$}
& \multicolumn{2}{c}{$^{16}{\rm O}^{5+}$}
& \multicolumn{2}{c}{$^{20}{\rm Ne}^{7+}$}
& \multicolumn{2}{c}{$^{24}{\rm Mg}^{9+}$}
\\ \hline
Dirac value (point nucleus)  &    1.&999 680 300      &    1.&999 431 380      &    1.&999 110 996      &    1.&998 718 893      \\
Finite nuclear size          &    0.&000 000 000      &    0.&000 000 000      &    0.&000 000 001      &    0.&000 000 001      \\
Interelectronic interaction  &    0.&000 130 758 (19) &    0.&000 176 658 (30) &    0.&000 222 628 (42) &    0.&000 268 703 (55) \\
QED, $\sim \alpha$           &    0.&002 323 017 (1)  &    0.&002 323 182 (1)  &    0.&002 323 405 (2)  &    0.&002 323 691 (2)  \\
QED, $\sim \alpha^2$         & $-$0.&000 003 515      & $-$0.&000 003 516      & $-$0.&000 003 516      & $-$0.&000 003 516      \\
Screened QED                 & $-$0.&000 000 085 (6)  & $-$0.&000 000 117 (12) & $-$0.&000 000 150 (21) & $-$0.&000 000 183 (32) \\
Nuclear recoil               &    0.&000 000 010      &    0.&000 000 017      &    0.&000 000 025      &    0.&000 000 032      \\
Total                        &    2.&002 130 485 (19) &    2.&001 927 604 (32) &    2.&001 653 389 (47) &    2.&001 307 619 (64) \\
\hline
\hline
& \multicolumn{2}{c}{$^{32}{\rm S}^{13+}$}
& \multicolumn{2}{c}{$^{40}{\rm Ar}^{15+}$}
& \multicolumn{2}{c}{$^{40}{\rm Ca}^{17+}$}
& \multicolumn{2}{c}{$^{52}{\rm Cr}^{21+}$}
\\ \hline
Dirac value (point nucleus)  &    1.&997 718 193      &    1.&997 108 781      &    1.&996 426 011      &    1.&994 838 064      \\
Finite nuclear size          &    0.&000 000 005      &    0.&000 000 009      &    0.&000 000 014      &    0.&000 000 035      \\
Interelectronic interaction  &    0.&000 361 24 (9)   &    0.&000 407 75 (12)  &    0.&000 454 45 (14)  &    0.&000 548 48 (21)  \\
QED, $\sim \alpha$           &    0.&002 324 470 (3)  &    0.&002 324 973 (5)  &    0.&002 325 555 (5)  &    0.&002 326 984 (5)  \\
QED, $\sim \alpha^2$         & $-$0.&000 003 518 (1)  & $-$0.&000 003 519 (1)  & $-$0.&000 003 520 (2)  & $-$0.&000 003 523 (6)  \\
Screened QED                 & $-$0.&000 000 25 (6)   & $-$0.&000 000 29 (8)   & $-$0.&000 000 33 (10)  & $-$0.&000 000 42 (15)  \\
Nuclear recoil               &    0.&000 000 046 (1)  &    0.&000 000 048 (1)  &    0.&000 000 061 (2)  &    0.&000 000 070 (4)  \\
Total                        &    2.&000 400 19 (11)  &    1.&999 837 75 (14)  &    1.&999 202 24 (17)  &    1.&997 709 69 (26)  \\
\hline
\hline
& \multicolumn{2}{c}{$^{74}{\rm Ge}^{29+}$}
& \multicolumn{2}{c}{$^{132}{\rm Xe}^{51+}$}
& \multicolumn{2}{c}{$^{208}{\rm Pb}^{79+}$}
& \multicolumn{2}{c}{$^{238}{\rm U}^{89+}$}
\\ \hline
Dirac value (point nucleus)  &    1.&990 752 307      &    1.&972 750 205      &    1.&932 002 904      &    1.&910 722 624 (1)  \\
Finite nuclear size          &    0.&000 000 162      &    0.&000 003 37 (1)   &    0.&000 078 58 (13)  &    0.&000 241 30 (43)  \\
Interelectronic interaction  &    0.&000 739 75 (37)  &    0.&001 299 4 (11)   &    0.&002 140 7 (27)   &    0.&002 501 4 (38)   \\
QED, $\sim \alpha$           &    0.&002 330 979 (6)  &    0.&002 351 91 (2)   &    0.&002 411 7 (1)    &    0.&002 446 3 (2)    \\
QED, $\sim \alpha^2$         & $-$0.&000 003 523 (24) & $-$0.&000 003 54 (13)  & $-$0.&000 003 6 (5)    & $-$0.&000 003 6 (8)    \\
Screened QED                 & $-$0.&000 000 45 (20)  & $-$0.&000 001 2 (4)    & $-$0.&000 003 5 (12)   & $-$0.&000 003 1 (15)   \\
Nuclear recoil               &    0.&000 000 092 (9)  &    0.&000 000 16 (6)   &    0.&000 000 25 (35)  &    0.&000 000 28 (69)  \\
Nuclear polarization         &      &                 &      &                 & $-$0.&000 000 04 (2)   & $-$0.&000 000 27 (14)  \\
Total                        &    1.&993 819 32 (42)  &    1.&976 400 3 (12)   &    1.&936 627 0 (30)   &    1.&915 904 9 (42)   \\
\hline
\hline
\end{tabular}
\end{table}
\end{document}